\documentclass[12pt]{article}
\usepackage{amssymb,amsmath,epsfig}

\begin{document}

\title{\bf Effects of Electromagnetic Field on Energy Density Inhomogeneity
in Self-Gravitating Fluids}
\author{M. Sharif \thanks {msharif.math@pu.edu.pk} and Neelum Bashir
\thanks {neelumkhan77@yahoo.com}\\
Department of Mathematics, University of the Punjab,\\
Quaid-e-Azam Campus, Lahore-54590, Pakistan.}

\date{}

\maketitle
\begin{abstract}
This paper is devoted to study the effects of electromagnetic field
on the energy density inhomogeneity in the relativistic
self-gravitating fluids for spherically symmetric spacetime. Two
important equations of the Weyl tensor are formulated which help to
analyze the energy density inhomogeneity in this scenario. We
investigate two types of fluids, i.e., non-dissipative and
dissipative. The non-dissipative fluid further includes dust,
locally isotropic, and locally anisotropic charged fluids. We
explore the effects of different factors on energy density
inhomogeneity in all these cases, in particular, the effect of
charge.
\end{abstract}

\section{Introduction}

Gravitational collapse emerges from a highly inhomogeneous initial
state which can be explained in terms of inhomogeneous energy
density distribution. Energy density inhomogeneity plays a vital
role in the collapse of self-gravitating fluid. Penrose \cite{1}
discussed the role of the Weyl tensor in the evolution of
self-gravitating system. He provided a simple relation between the
Weyl tensor and energy density to investigate the gravitational
arrow of time. The Weyl tensor may be represented exclusively in
terms of energy density and local anisotropy of pressure, which
affects the fate of gravitational collapse.

There is a large body of literature available \cite{13}-\cite{7}
which indicate the importance of energy density inhomogeneity in
self-gravitating fluid. Joshi and Dwivedi \cite{13} analyzed the
Tolman-Bondi model to examine the nature and occurence of naked
singularities for inhomogeneous gravitational collapse. Triginer and
Pavon \cite{2} discussed heat transport equation in an inhomogeneous
spherically symmetric universe. Mena and Tavokol \cite{12}
investigated how it is important in the dust collapse. Herrera et
al. \cite{6} derived a relation for the active gravitational mass of
collapsing fluid distribution to explain the effects of density
inhomogeneity and local anisotropy on spherical collapse. The same
authors \cite{7} also studied the behavior of locally anisotropic,
self-gravitating spherical symmetric dissipative fluid with the Weyl
tensor and density inhomogeneity.

The study of self-gravitating spherically symmetric charged fluid
distribution has been the subject of interest for many people.
Rosseland \cite{15} and Eddington \cite{16} started the study of
self-gravitating spherically symmetric charged fluid distribution.
Since then, a lot of work has been done to investigate the effects
of electric charge on the structure and evolution of
self-gravitating system \cite{22}-\cite{17}. Di Prisco et al.
\cite{38} investigated the non-adiabatic charged spherically
symmetric gravitational collapse as well as the energy density
inhomogeneity. Sharif and his collaborators \cite{33} derived
dynamical as well as transport equations of matter dissipating in
the form of shear viscosity to see the effect of charge on
gravitational collapse.

The gravitational collapse is highly dissipative process
\cite{25,26}, hence its effects are much important in the study of
collapse. Misner \cite{27} explained the dissipative relativistic
gravitational collapse by using the streaming out approximation.
Two possible approximations (diffusion and streaming out) are
usually considered in dissipative process. Diffusion approximation
is the approximation for which energy flux of radiation (like
thermal conduction) is proportional to the gradient of
temperature. Israel and Stewart \cite{28} formulated transport
equation required for diffusion approximation . Lattimer \cite{29}
explained that during emission process, the role of radiation
transport is closer to diffusion approximation than the streaming
out approximation. Recently, Herrera \cite{30} explored energy
density inhomogeneity in self-gravitating fluid as well as its
stability.

In this paper, we extend this work to see the effects of charge on
energy density inhomogeneity due to fluid distribution. We shall
discuss the effects of electric charge on the relationship between
the Weyl tensor and energy density inhomogeneity. The outline of the
paper is as follows. In the next section, we review the relevant
kinematics and Einstein-Maxwell field equations for a spherically
symmetric distribution of collapsing charged fluid. Section
\textbf{3} is devoted to dynamical as well as transport equations
and two equations of the Weyl tensor. In section \textbf{4},
different aspects of fluid are considered to explore energy density
inhomogeneity. The last section contains the conclusion of the
results.

\section{Spacetime and Matter Distribution}

We take spherically symmetric distribution of dissipative
collapsing charged fluid bounded by a spherical surface $\Sigma$
in the comoving coordinates as
\begin{equation}\label{1}
ds^{2}=-A^2dt^{2}+B^2dr^{2}+C^2(d\theta^2+\sin^2\theta d\phi^2),
\end{equation}
where $A,~B$ and $C$ are functions of $t$ and $r$. Matter under
consideration is anisotropic fluid suffering dissipation in the
form of heat, i.e.,
\begin{eqnarray}\label{2}
T^{(m)}_{\alpha\beta}=(\mu+P_\perp)V_\alpha V_\beta +P_\perp
g_{\alpha\beta} +(P_r-P_\perp)\chi_\alpha \chi_\beta +q_\alpha
V_\beta +V_\alpha q_\beta +\epsilon l_\alpha l_\beta ,
\end{eqnarray}
where $\mu,~P_r,~P_\perp,~V^\alpha,~\chi^\alpha,~q^\alpha,~l^\alpha$
and $\epsilon$ represent the energy density, the radial pressure,
the tangential pressure, the four velocity of the fluid, a unit
four-vector along the radial direction, the heat flux, a radial null
four-vector and the energy density of null fluid describing
dissipation in the free streaming approximation respectively. These
quantities satisfy
\begin{eqnarray*}\nonumber
V^\alpha V_\beta=-1,\quad V^\alpha q_\beta=0,\quad
\chi^\alpha\chi_\beta=1,\quad l^\alpha V_\alpha=-1,\quad \chi^\alpha
V_\alpha=0,\quad l^\alpha l_\alpha=0.
\end{eqnarray*}
Also, we define
\begin{eqnarray*}\label{4}
V^\alpha =A^{-1} \delta^\alpha_0,\quad q^\alpha=q B^{-1}
\delta^\alpha_1,\quad l^\alpha =A^{-1} \delta^\alpha_0 + B^{-1}
\delta^\alpha_1,\quad \chi^\alpha =B^{-1} \delta^\alpha_1.
\end{eqnarray*}
The shear tensor is given by
\begin{equation*}\label{5}
\sigma_{\alpha\beta}=V_{(\alpha
;\beta)}+a_{(\alpha}V_{\beta)}-\frac{1}{3} \Theta h_{\alpha\beta},
\end{equation*}
where $h_{\alpha\beta}=g_{\alpha\beta}+ V_\alpha V_\beta$. The four
acceleration $a_\alpha$ and the expansion $\Theta$ are
\begin{equation*}\label{6}
a_\alpha=V_{\alpha;\beta} V^\beta,\quad
\Theta=V^\alpha~_{;\alpha},
\end{equation*}
which yield
\begin{equation}\label{7}
a_1=\frac{A'}{A},\quad a^2=a^\alpha
a_\alpha=\left(\frac{A'}{AB}\right)^2,\quad \Theta=\frac{1}{A}
\left(\frac{\dot{B}}{B}+2 \frac{\dot{C}}{C}\right).
\end{equation}
The non-zero components of the shear tensor will be
\begin{equation*}\label{9}
\sigma_{1 1}=\frac{2}{3} B^2 \sigma,\quad \sigma_{2
2}=\frac{\sigma_{3 3}}{\sin^2\theta} =-\frac{1}{3} C^2 \sigma,
\end{equation*}
with its scalar
\begin{equation}\label{10}
\sigma^{\alpha \beta}\sigma_{\alpha
\beta}=\frac{2}{3}\sigma^2,\quad \sigma=\frac{1}{A}
\left(\frac{\dot{B}}{B}-\frac{\dot{C}}{C}\right).
\end{equation}

The electromagnetic energy-momentum tensor is
\begin{equation}\label{13}
T^{(em)}_{\alpha \beta}=\frac{1}{4 \pi}(F^\gamma_\alpha
F_{\beta\gamma} -\frac{1}{4}F^{\gamma\delta}F_{\gamma\delta}
g_{\alpha \beta}),
\end{equation}
where $F_{\alpha\beta}$ is the electromagnetic field tensor. In
terms of four-vector formulation, the Maxwell field equations can
be written in the form
\begin{equation}\label{14}
F_{\alpha \beta}=\phi_{\beta,\alpha}-\phi_{\alpha,\beta},\quad
F^{\alpha \beta}_{;\beta}=\mu_{0}J^\alpha,
\end{equation}
where $\phi_\alpha,~J^\alpha$ and $\mu_{0}$ represent the four
potential, four current and magnetic permeability, respectively.
Since charge is considered at rest in comoving coordinates, so
$J^\alpha$ and $\phi_\alpha$ become
\begin{equation*}\label{16}
\phi_\alpha=\phi\delta^{0}_\alpha ,\quad J^\alpha=\rho V^\alpha.
\end{equation*}
Here the charge density $\rho$ and electric scalar potential
$\phi$ are both functions of $t$ and $r$. The charge conservation
equation gives
\begin{equation}\label{17}
s(r)=4\pi\int^r_0\rho BC^2dr,
\end{equation}
where $s(r)$ is the electric charge interior to radius $r$. For
metric (\ref{1}), the Maxwell equations turn out to be
\begin{eqnarray}\label{18}
\phi''-\left(\frac{A'}{A}+\frac{B'}{B}-2\frac{C'}{C}\right)
\phi'&=&4\pi\rho A B^2 ,\\\label{19}
\dot{\phi'}-\left(\frac{\dot{A}}{A}\frac{\dot{B}}{B}-2\frac{\dot{C}}{C}\right)
\phi'&=&0.
\end{eqnarray}
Equation (\ref{18}) yields
\begin{equation}\label{20}
\phi'=\frac{s A B}{C^2},
\end{equation}
which satisfies Eq.(\ref{19}). The field equations for charged
dissipative fluid, $G_{\alpha \beta}=8
\pi(T^{(m)}_{\alpha\beta}+T^{(em)}_{\alpha\beta})$, are \cite{38}
\begin{eqnarray}\nonumber
8\pi(T^{(m)}_{00}+T^{(em)}_{00})&=&8\pi(\mu+\epsilon)A^2+\frac{(sA)^2}{C^4}\\\nonumber
&=&\left(2\frac{\dot{B}}{B}+\frac{\dot{C}}{C}\right)
\frac{\dot{C}}{C}-\left(\frac{A}{B}\right)^2\\\label{22}
&\times&\left[2\frac{C''}{C}+\left(\frac{C'}{C}\right)^2-2\frac{B'
C'}{BC}-\left(\frac{B}{C}\right)^2\right], \\\nonumber 8
\pi(T^{(m)}_{01}+T^{(em)}_{01})&=&-8\pi(q+\epsilon)AB\\\label{23}
&=&-2\left(\frac{\dot{C'}}{C}-\frac{\dot{B}C'}{B
C}-\frac{\dot{C}A'}{CA}\right),\\\nonumber
8\pi(T^{(m)}_{11}+T^{(em)}_{11})&=&8\pi\left(P_r+\epsilon
\right)B^2-\frac{s^2B^2}{C^4}\\\nonumber
&=&-\left(\frac{B}{A}\right)^2\left[2\frac{\ddot{C}}{C}
-\left(2\frac{\dot{A}}{A}-\frac{\dot{C}}{C}\right)
\frac{\dot{C}}{C}\right]\\\label{24}
&+&\left(2\frac{A'}{A}+\frac{C'}{C}\right)\frac{C'}{C}-\left(\frac{B}{C}\right)^2,
\end{eqnarray}
\begin{eqnarray}\nonumber
8\pi(T^{(m)}_{22}+T^{(em)}_{22})&=&\frac{8\pi}{\sin^2\theta}(T^{(m)}_{33}+T^{(em)}_{33})=8\pi
P_\perp C^2+\frac{s^2}{C^2}\\\nonumber
&=&\left(\frac{C}{A}\right)^2\left[\frac{\dot{A}}{A}\left(\frac{\dot{B}}{B}
+\frac{\dot{C}}{C}\right)-\frac{\ddot{B}}{B}-\frac{\ddot{C}}{C}
-\frac{\dot{B}\dot{C}}{BC}\right]\\\label{25}
&+&\left(\frac{C}{B}\right)^2\left[\frac{A''}{A}+\frac{C''}{C}-\frac{A'
B'}{AB}+\left(\frac{A'}{A}-\frac{B'}{B}\right)\frac{C'}{C}\right].
\end{eqnarray}

\section{Dynamical and Transport Equations}

Using the Misner-Sharp definition \cite{2222}, the mass function
in the presence of charge is given by
\begin{equation}\label{26}
m=\frac{C^3}{2}R_{23}~^{23}+\frac{s^2}{2C}=
\frac{C}{2}\left[\left(\frac{\dot{C}}{A}\right)^2
-\left(\frac{C'}{B}\right)^2+1\right]+\frac{s^2}{2C}.
\end{equation}
The proper time and radial derivatives are defined by \cite{38}
\begin{equation}\label{27}
D_T=\frac{1}{A}\frac{\partial}{\partial t},\quad
D_R=\frac{1}{C'}\frac{\partial}{\partial r},
\end{equation}
where $R$ stands for proper areal radius. Using this definition,
the velocity of the collapsing fluid is
$U=D_TC=\frac{\dot{C}}{A}<0$. Thus we can write  Eq.(\ref{26}) as
\begin{equation}\label{30}
\textit{E}=\frac{C'}{B}=\sqrt{1+U^2-\frac{2m}{C}+\left(\frac{s}{C}\right)^2}.
\end{equation}
Define
$\tilde{\mu}=\mu+\epsilon,~\tilde{P_r}=P_r+\epsilon,~\tilde{q}=q+\epsilon$.
The rate of change of mass is
\begin{equation}\label{31}
D_T m=-4\pi\left[\tilde{P_r}U+\tilde{q}\textit{E}\right]C^2,\quad
D_R
m=4\pi\left[\tilde{\mu}+\tilde{q}\frac{U}{\textit{E}}\right]C^2+\frac{s}{C}D_Rs.
\end{equation}
This shows how mass is affected by different quantities.
Integration of the second equation yields
\begin{equation}\label{33}
m=\int_0^r\left[4\pi(\tilde{\mu}+\tilde{q}\frac{U}{\textit{E}})C^2C'+\frac{ss'}{C}\right]dr.
\end{equation}
The contracted Bianchi identities,
$(T^{(m)^{\alpha\beta}}+T^{(em)^{\alpha\beta}})_{;\beta}=0$, yield
\begin{eqnarray}\nonumber
(T^{(m)^{\alpha\beta}}+T^{(em)^{\alpha\beta}})_{;\beta} V_\alpha
&=&\dot{\tilde{\mu}}+(\tilde{\mu}+\tilde{P_r})\frac{\dot{B}}{B}
+2(\tilde{\mu}+\tilde{P_\perp})\frac{\dot{C}}{C}\\\label{37}&+&\frac{\tilde{q'A}}{B}
+2\tilde{q}\frac{(AC)'}{BC},\\\nonumber
(T^{(m)^{\alpha\beta}}+T^{(em)^{\alpha\beta}})_{;\beta}
\chi_\alpha
&=&\dot{\tilde{q}}+\frac{\tilde{(P_r)'}A}{B}+2\tilde{q}
\left(\frac{\dot{B}}{B}+\frac{\dot{C}}{C}\right)
\\\label{38}&+&(\tilde{\mu}+\tilde{P_r})\frac{A'}{B}+2\Pi
\frac{AC'}{BC}-\frac{ss'A}{4\pi BC^4}=0,
\end{eqnarray}
where $\Pi=\tilde{P_r}-P_{\bot}$. These are called dynamical
equations.

The Weyl tensor is defined by
\begin{eqnarray}\nonumber
C^\rho_{\alpha\beta\mu}&=&R^\rho_{\alpha\beta\mu}-\frac{1}{2}R^\rho_\beta
g_{\alpha\mu}+\frac{1}{2}R_{\alpha\beta}
\delta^\rho_\mu-\frac{1}{2}R_{\alpha\mu}
\delta^\rho_\beta+\frac{1}{2}R^\rho_\mu g_{\alpha\beta}\\\label{39}
&+&\frac{1}{6}R(\delta^\rho_\beta g_{\alpha\beta}-g_{\alpha\beta}
\delta^\rho_\mu).
\end{eqnarray}
The electric part of the Weyl tensor,
$E_{\alpha\beta}=C_{\alpha\mu\beta\nu}V^\mu V^\nu$, gives
\begin{equation*}\label{41}
E_{11}=\frac{2}{3}B^2\varepsilon ,\quad
E_{22}=-\frac{1}{3}C^2\varepsilon ,\quad
E_{33}=E_{22}\sin^2\theta,
\end{equation*}
where
\begin{eqnarray}\nonumber
\varepsilon&=&\frac{1}{2A^2}\left[\frac{\ddot{C}}{C}-\frac{\ddot{B}}{B}
-\left(\frac{\dot{C}}{C}-\frac{\dot{B}}{B}\right)
\left(\frac{\dot{A}}{A}+\frac{\dot{C}}{C}\right)\right]
\\\label{44}&+&\frac{1}{2B^2}\left[\frac{A''}{A}-\frac{C''}{C}
+\left(\frac{B'}{B}+\frac{C'}{C}\right)\left(\frac{C'}{C}
-\frac{A'}{A}\right)\right]-\frac{1}{2C^2},
\end{eqnarray}
while its magnetic part vanishes due to spherical symmetry. We may
also write $E_{\alpha\beta}$ as
\begin{equation}\label{45}
E_{\alpha\beta}=\varepsilon(\chi_\alpha\chi_\beta-\frac{1}{3}h_{\alpha\beta}).
\end{equation}
Solving Eqs.(\ref{22})-(\ref{24}) with Eqs.(\ref{26}) and
(\ref{44}), we get
\begin{equation}\label{46}
\frac{3m}{C^3}-\frac{2s^2}{C^4}=4\pi(\tilde{\mu}-\Pi)-\varepsilon.
\end{equation}
Using Eqs.(\ref{31}) and (\ref{44}), we can write
\begin{eqnarray}\label{48}
[{\varepsilon-4\pi(\tilde{\mu}-\Pi)}\dot{]}
&=&3\frac{\dot{C}}{C}\left[4\pi(\tilde{\mu}
+P_\perp)-\frac{2s^2}{3C^4}-\varepsilon\right]
+12\pi\tilde{q}\frac{AC'}{BC},\\\label{49}
[{\varepsilon-4\pi(\tilde{\mu}-\Pi)}]'
&=&-3\frac{C'}{C}\left[\varepsilon+4\pi\Pi+\frac{2s^2}{3C^4}\right]
-12\pi\tilde{q}\frac{\dot{C}B}{AC}+\frac{ss'}{C^4}.
\end{eqnarray}
These equations yield a relationship between the Weyl tensor,
energy density and charge that help to discuss energy density
inhomogeneities given in the next section.

The corresponding transport equation for heat flux derived from
Muller-Israel-Stewart theory \cite{33, 222} is given as
\begin{equation}\label{50}
\tau h^{\alpha\beta} V^\gamma q_{\beta;\gamma}+q^\alpha= -\kappa
h^{\alpha\beta}(T_{,\beta}+Ta_\beta)-\frac{1}{2} \kappa
T^2\left(\frac{\tau V^\beta}{\kappa T^2}\right)_{;\beta} q^\alpha,
\end{equation}
where $\kappa,~T$ and $\tau$ represent thermal conductivity,
temperature and relaxation time, respectively. Due to symmetry, the
above equation has only one independent component given by
\begin{equation}\label{51}
\tau \dot{q}=-\frac{1}{2}\kappa qT^2\left(\frac{\tau}{\kappa T^2
}\right)-\frac{1}{2}\tau q\Theta A-\frac{\kappa}{B}(TA)'-qA.
\end{equation}
For $\tau=0$, Eckart-Landau \cite{37} equation is recovered. In case
of truncated version of the theory, the last term in Eq.(\ref{50})
is removed and finally, we get
\begin{equation}\label{52}
\tau \dot{q}=-\frac{\kappa}{B} (TA)'-qA,
\end{equation}
which is the same as given in \cite{30}. This shows that the
electromagnetic field does not affect the heat transport equation.

\section{Energy Density Inhomogeneity}

In this section, we consider different aspects of fluid distribution
that are responsible for energy density inhomogeneity .

\subsection{Non-dissipative Charged Dust}

Firstly, the case of non-dissipative charged dust is taken, i.e.,
$q=P_r=P_\perp=\epsilon=0$. Since the fluid is moving along
geodesic (i.e., $a^{\alpha}=0=a)$, so Eq.(\ref{7}) leads to $A=1$.
Consequently, Eqs.(\ref{48}) and (\ref{49}) reduce to
\begin{eqnarray}\label{53}
&&(\varepsilon-4\pi\mu)^.+3\frac{\dot{C}}{C}
\left(\varepsilon-4\pi\mu+\frac{2s^2}{3C^4}\right)=0,\\\label{54}
&&(\varepsilon-4\pi\mu)'=-3\frac{C'}{C}\varepsilon-2\frac{s^2 C'
}{C^5}+\frac{ss'}{C^4}.
\end{eqnarray}
For $s=0$, this case reduces to the uncharged case \cite{30}.
Equation (\ref{54}) implies that for $\varepsilon=0=s$, we get
$\mu'=0$. This shows that energy density inhomogeneity depends not
only on the Weyl tensor but also on charge $s$. For
$\varepsilon=0=\mu'$ (i.e., when spacetime is conformally flat and
energy density vanishes), it follows from Eq.(\ref{54}) that
$C^2(t,r)= s(r)\phi(t)$, where $\phi(t)$ is an arbitrary function.
When $\mu'=0$, we obtain
\begin{equation}\label{56}
\varepsilon=\frac{1}{C^3}{\int^r_0(\frac{ss'}{C}-\frac{2s^2
C'}{C^2})dr},
\end{equation}
where integration function is chosen such that $\varepsilon(t,0)=0$.
This equation shows that homogeneity in energy density implies the
existence of $\varepsilon$ in the presence of electromagnetic field.

Next, we make use of Eqs.(\ref{10}) and (\ref{37}) in (\ref{53})
so that
\begin{equation}\label{58}
\varepsilon=-\frac{4\pi}{C^3}{\int^t_0\mu\sigma
C^3dt+\frac{2s^2}{C^4}}.
\end{equation}
Here integration function is chosen as $\varepsilon(0,r)=0$. With
the help of Raychaudhuri equation and the field equations, we get
an evolution equation for shear (see \cite{31} for detail]) given
by
\begin{equation}\label{59}
\dot{\sigma}+\frac{\sigma^2}{3}+2\frac{\Theta\sigma}{3}=
-\varepsilon.
\end{equation}
Equations (\ref{58}) and (\ref{59}) show that conformal flatness
and shearfree conditions do not imply each other. In this case,
the shearfree implies vanishing of the Weyl tensor but converse is
not true.

\subsection{Locally Isotropic Non-dissipative Charged Fluid}

This case corresponds to a non-dissipative charged isotropic
fluid, i.e., $\Pi=q=\epsilon=0,~P_r=P_\perp=P$. Equations
(\ref{48}) and (\ref{49}) reduce to
\begin{eqnarray}\label{60}
(\varepsilon-4\pi\mu)^.+3\frac{\dot{C}}{C}
\left(\varepsilon-4\pi(\mu+P)+\frac{2s^2}{3C^4}\right)=0,\\\label{61}
(\varepsilon-4\pi\mu)'=
-3\frac{C'}{C}\varepsilon-2\frac{s^2C'}{C^5}+\frac{ss'}{C^4}.
\end{eqnarray}
Equation (\ref{61}) is exactly the same as (\ref{54}) and hence
the same behavior for $\varepsilon=q=0,~\varepsilon=\mu'=0$ and
$~\mu'=0 $. Using Eqs.(\ref{10}) and (\ref{37}) in (\ref{60}), we
have
\begin{equation}\label{63}
\varepsilon=-\frac{4\pi}{C^3}\int^t_0[(\mu+P)A\sigma
C^3]dt+\frac{2s^2}{C^4},
\end{equation}
where $\varepsilon(0,r)=0$. For non-dissipative locally isotropic
fluid (using Raychaudhuri equation as well field equations as in the
non-dissipative charged case), the evolution equation of the shear
takes the form
\begin{equation}\label{64}
\varepsilon=\frac{a'}{B}-\frac{\dot{\sigma}}{A}+a^2-\frac{\sigma^2}{3}
-\frac{2}{3}\Theta\sigma-a\frac{C'}{BC}.
\end{equation}
It is important to mention here that neither the vanishing of the
Weyl tensor implies shearfree nor the shearfree condition
corresponds to conformally flatness (i.e., the Weyl tensor
disappears). If we assume the fluid to be shearfree, then
Eq.(\ref{63}) yields
\begin{equation}\label{65}
\varepsilon=\frac{2s^2}{C^4}.
\end{equation}
This shows that under the effect of electromagnetic field for
initial homogeneous configuration, i.e, $\varepsilon(0,r)=0$, the
Weyl tensor does not disappear at any time $t$.

\subsection{Locally Anisotropic Non-dissipative Charged Fluid}

In this case, we consider the role of pressure anisotropy with
$q=\epsilon=0$ but $\Pi\neq0$. Under these conditions,
Eqs.(\ref{48}) and (\ref{49}) become
\begin{eqnarray}\label{66}
(\varepsilon-4\pi\mu+4\pi\Pi)^.+3\frac{\dot{C}}{C}
\left(\varepsilon-4\pi(\mu+P_\perp)+\frac{2s^2}{3C^4}\right)=0,\\\label{67}
(\varepsilon-4\pi\mu+4\pi\Pi)'+3\frac{C'}{C}\left(\varepsilon+4\pi\Pi
+\frac{2s^2}{3C^4}\right)-\frac{ss'}{C^4}=0.
\end{eqnarray}
In the case of locally anisotropic fluid, the quantity
$\varepsilon+4\pi\Pi$ plays an important role instead of the Weyl
tensor. The existence of the energy density inhomogeneity depends on
$\varepsilon+4\pi\Pi$ as well as charge unlike previous cases. From
Eq.(\ref{67}), it follows that for $\varepsilon+4\pi\Pi=0=s$, we get
$\mu'=0$. When we take $\varepsilon+4\pi\Pi=0=\mu'$, then
Eq.(\ref{67}) reduces to $C^2(t,r)=s(r)\phi(t)$ while for $\mu'=0$,
this yields
\begin{equation}\label{68}
\left(-\epsilon-4\pi\Pi+\frac{s^2}{C^4}\right)'+3\frac{(-\epsilon-4\pi\Pi
+\frac{s^2}{C^4})C'}{C}= \frac{ss'}{C^4}+\frac{s^2C'}{C^5}.
\end{equation}
Using Eqs.(\ref{10}) and (\ref{37}) in (\ref{66}), it follows that
\begin{eqnarray}\nonumber
&&\left(-\epsilon-4\pi\Pi+\frac{s^2}{C^4}\right)^.
+3\frac{(-\epsilon-4\pi\Pi+\frac{s^2}{C^4})
\dot{C}}{C}\\\label{69} &&=4\pi(\mu+P_r)\sigma A
-\frac{8\pi\Pi\dot{C}}{C}+\frac{s^2\dot{C}}{C^5}.
\end{eqnarray}
These two equations represent evolution equations for the quantity
$-(\varepsilon+4\pi\Pi)+\frac{s^2}{C^4}$ which represents one of
the structure scalars $X_{TF}$.

The tensor $X_{\alpha\beta}$ is defined as
\begin{equation}\label{71}
X_{\alpha\beta}=^* R^*_{\alpha\gamma\beta\delta}=
\frac{1}{2}\eta_{\alpha\gamma}~^{\varepsilon\rho}
R^*_{\varepsilon\rho\beta\delta}V^\gamma V^\delta,
\end{equation}
where $R^*_{\alpha\beta\gamma\delta}
=\frac{1}{2}\eta_{\varepsilon\rho\gamma\delta}
R_{\alpha\beta}~^{\varepsilon\rho}$ and
$\eta_{\varepsilon\rho\gamma\delta}$ denote the Levi-Civita
tensor. Tensor $X_{\alpha\beta}$ can be expressed through its
trace and trace free part as
\begin{equation}\label{72}
X_{\alpha\beta}=\frac{1}{3}X_T
h_{\alpha\beta}+X_{TF}(\chi_\alpha\chi_\beta-\frac{1}{3}h_{\alpha\beta}).
\end{equation}
One can find $X_{TF}$ with the help of the field equations,
Eq.(\ref{72}) and (\ref{39})-(\ref{45}) ( see for detail
\cite{444}) as
\begin{equation}\label{73}
X_{TF}=-\epsilon-4\pi\Pi+\frac{s^2}{C^4}.
\end{equation}
Thus the evolution equations (\ref{68}) and (\ref{69}) in terms of
$X_{TF}$ turn out to be
\begin{eqnarray}\nonumber
&&X_{TF}'+3\frac{X_{TF}C'}{C}= \frac{s^2 C'}{C^5}+\frac{ss'}{C^4},
\\\nonumber
&&\dot{X}_{TF}+3\frac{X_{TF}\dot{C}}{C}= 4\pi(\mu+P_r)\sigma A
-\frac{8\pi\Pi\dot{C}}{C}+\frac{s^2\dot{C}}{C^5}.
\end{eqnarray}
The corresponding solutions will be
\begin{eqnarray}\label{76}
&&X_{TF}=\frac{1}{C^3}{\int^r_0 (-8\pi\Pi C'C^2+\frac{s^2
C'}{C^2}+\frac{ss'}{C})dr},\label{77}\\
&&X_{TF}=\frac{-4\pi}{C^3}{\int^t_0
[2\pi\dot{C}-(\mu+P_r)AC\sigma]C^2dt}-\frac{s^2}{C^4}.
\end{eqnarray}
The last equation represents that the anisotropy of pressure and
electromagnetic field affect the initial state of the energy
density.

\subsection{Dissipative Geodesic Charged Dust}

Here we consider the case of dissipative geodesic dust to explore
the effect of dissipation on energy density inhomogeneity. For this
purpose, we assume that dissipative dust is moving along geodesic,
i.e., $P_r=P_\perp=0$ and $A=1$. The corresponding equations take
the form
\begin{eqnarray}\label{78}
&&(\varepsilon-4\pi\tilde{\mu})^.+3\frac{\dot{C}}{C}
\left(\varepsilon-4\pi\tilde{\mu}+\frac{2s^2}{3C^4}\right)
-\frac{12\pi\tilde{q}C'}{BC}=0,\\\label{79}
&&(\varepsilon-4\pi\tilde{\mu})'=
-3\frac{C'}{C}\varepsilon-2\frac{s^2C'}{C^5}
-\frac{12\pi\tilde{q}\dot{C}B}{C}+\frac{ss'}{C^4}.
\end{eqnarray}
Equation (\ref{79}) leads to a quantity $\Psi$
\begin{equation}\label{80}
\Psi=\varepsilon+\frac{12\pi}{C^3}{\int^r_0}\tilde{q}\dot{C}BC^2dr.
\end{equation}
Equation (\ref{79}) implies that for $\Psi=s=0$, we get
$\tilde{\mu'}=0$ similar to the uncharged case. For
$\Psi=\tilde{\mu'}=0$, we obtain a relation between $s$ and $C$
from (\ref{79}) as discussed in the previous cases.

Taking $\tilde{\mu'}=0$ in Eq.(\ref{79}), it follows that
\begin{equation}\label{81}
\Psi=\frac{1}{C^3}{\int^r_0 (\frac{ss'}{C}-2\frac{s^2 C'}{C})dr}.
\end{equation}
Using Eqs.(\ref{78}), (\ref{10}) and (\ref{37}), we obtain the
evolution equation of $\Psi$
\begin{equation}\label{82}
\dot{\Psi}-\frac{\dot{\Omega}}{C^3}=-4\pi\tilde{\mu}\sigma
-\frac{4\pi\tilde{q'}}{B}+\frac{4\pi\tilde{q}C'}{BC}+\frac{2s^2\dot{C}}{C^5},
\end{equation}
where
\begin{equation}\label{83}
\Omega=12\pi\int^r_0\tilde{q}\dot{C}BC^2dr.
\end{equation}
The general solution of Eq.(\ref{82}) is
\begin{equation}\label{84}
\Psi=\frac{1}{C^3}{\int^t_0(\dot{\Omega}-4\pi\sigma\tilde{\mu}C^3
-\frac{4\pi\tilde{q}'C^3}{B}+\frac{4\pi\tilde{q}C^2C'}{B}+\frac{2s^2\dot{C}}{C^2})dt}.
\end{equation}
This shows that charged dust density inhomogeneity depends on
dissipative terms, shear and the electromagnetic field, for the
initially homogeneous configuration. To investigate the effect of
these factors on evolution of $\Psi$, we consider shearfree case
for which Eq.(\ref{10}) implies that $C=Br$. Consequently,
Eq.(\ref{83}) yields
\begin{equation}\label{85}
\Omega=12\pi\int^r_0\frac{\tilde{q}\dot{C}C^3}{r}dr.
\end{equation}
The corresponding evolution equation is
\begin{equation}\label{86}
\Psi=\frac{1}{C^3}{\int^t_0(\dot{\Omega}-4\pi\tilde{q}'C^2
r+4\pi\tilde{q}CrC'+\frac{2s^2\dot{C}}{C^2})dt}.
\end{equation}

Finally, we discuss the relaxation effects in evolution of $\Psi$.
The diffusion approximation $(\epsilon=0)$ implies
$\tilde{q}=q,~\tilde{\mu}=\mu$. Thus from Eq.(\ref{38}), it
follows that
\begin{equation}\label{87}
\dot{q}=-\frac{4q}{3}\Theta+\frac{2ss'}{4\pi BC^4}.
\end{equation}
Combining the above equation with transport equation (\ref{54}), we
have
\begin{equation}\label{88}
q=-\frac{\kappa
T'}{B(1-\frac{4}{3}\Theta\tau)}-\frac{2ss'\tau}{4\pi
BC^4(1-\frac{4}{3}\Theta\tau)}.
\end{equation}
For the shearfree case, this reduces to
\begin{equation}\label{89}
q=-\frac{\kappa
rT'}{C(1-\frac{4}{3}\Theta\tau)}-\frac{2ss'r\tau}{4\pi
C^5(1-\frac{4}{3}\Theta\tau)}.
\end{equation}
When we insert this value of $q$ in Eq.(\ref{86}), we obtain $\Psi$
in terms of relaxation time $\tau$ which help to analyze relaxation
effects in terms of electromagnetic field.

\section{Conclusion}

In this paper, we have studied how charge affects the energy density
inhomogeneity and stability of the conformal flatness. For this
purpose, the evolution equations of the Weyl tensor are formulated.
We have investigated different aspects of fluid distribution
responsible for energy density inhomogeneity. The main results are
summarized as follows:
\begin{itemize}
\item In the case of non-dissipative charged dust and isotropic
fluid, we have found that when we take electromagnetic field
contribution, the energy density inhomogeneity is not controlled by
the Weyl tensor alone but it also depends on charge. It is worth
mentioning here that charge affects the conformal flat condition.
Here the vanishing of the Weyl tensor implies that shear depends on
charge. For shearfree fluid, the Weyl tensor disappears but this is
not true for isotropic fluid. In the absence of charge, our result
reduce to \cite{30}.
\item For anisotropic fluid, it is found that the shear inhomogeneity
also depends on charge in addition to anisotropy of pressure. A
specific quantity associated with energy density inhomogeneity is
identified as one of the structure scalar.
\item Finally, we have investigated the effect of charge on dissipative geodesic dust.
The effect of charge and different factors on the evolution of
$\Psi$ (a quantity that determines the existence of energy density
inhomogeneity) are explored. The relaxation effects in the evolution
of $\Psi$ with the inclusion of charge are also indicated in this
case.
\end{itemize}

\end{document}